# THERMODYNAMIC AND QUANTUM THERMODYNAMIC ANSWERS TO EINSTEIN'S CONCERNS ABOUT BROWNIAN MOVEMENT

Elias P. Gyftopoulos

On the occasion of the 100th anniversary of the beginning of the revolutionary contributions to physics by the "Person of the 20th Century", I am happy to respond to a question posed by him in 1905. In his articles "Investigations on the Brownian movement" Einstein said: "In this paper it will be shown that according to the molecular-kinetic theory of heat, bodies of microscopically-visible size suspended in a liquid will perform movements of such magnitude that they can be easily observed in a microscope, on account of the molecular motions of heat. It is possible that the movements to be discussed here are identical with the so called "Brownian molecular motion"; however, the information available to me regarding the latter is so lacking in precision, that I can form no judgment in the matter." And then he continues: "If the movement discussed here can actually be observed (together with the laws relating to it that one would expect to find), then classical thermodynamics can no longer be looked upon as applicable with precision to bodies even of dimensions distinguishable in a microscope; an exact determination of actual atomic dimensions is then possible. On the other hand, had the prediction of this movement proved to be incorrect, a weighty argument would be provided against the molecular-kinetic conception of heat." In this article I provide incontrovertible evidence against the molecular-kinetic conception of heat, and a regularization of the Brownian movement that differs from all the statistical procedures and/or analyses that exist in the archival literature to date. The regularization is based on either of two distinct but intimately interrelated revolutionary conceptions (in the sense of T. S. Kuhn) of thermodynamics by a group of faculty and students at the Massachusetts Institute of Technology over the past three decades. One is purely thermodynamic without any statistics of either statistical mechanics or probabilities of

**ELIAS P. GYFTOPOULOS** *is a Ford Professor Emeritus of the Departments of Mechanical and Nuclear Engineering at the Massachusetts Institute of Technology in Cambridge, Massachusetts.*

conventional quantum mechanics. The other is quantum mechanical but without statistical probabilities of statistical quantum mechanics. In the thermodynamic exposition we prove that entropy is an inherent – intrinsic – property of the constituents of a system in any state be it thermodynamic – stable – equilibrium or not thermodynamic equilibrium, and its analytic expression must satisfy eight criteria. In the unified theory, we prove that the quantum probabilities are described only by a density operator $\rho \geq \rho^2$ that can be represented by a homogeneous ensemble. In a homogeneous ensemble every member of the ensemble is characterized by the same $\rho$ as the whole ensemble or, equivalently, any conceivable subensemble, and more importantly for the purposes of this article, we show that in a state of thermodynamic equilibrium the velocity of each particle of each constituent (not the average of the velocities of many particles) is equal to zero, that is, nothing moves. In view of this result, one might think that we conclude that there is no Brownian movement, a conclusion contrary to the overwhelming and long lasting experimental evidence. But we will see that this interpretation of our conclusion is unwarranted because we prove that Brownian movement reflects other phenomena than motions of particles in the sense of the word "motion".

## Introduction

Brownian movement has a history of more than two centuries. Studies that appeared in the scientific archival literature in the period until 1920 are listed by Einstein in his booklet on "Investigations on the theory of the Brownian movement" [1]. Articles that were written until 1970 are presented in Refs. [2] and [3]. All these studies and articles are based on a variety of *statistical interpretations* of thermodynamics, and on the conception of the molecular-kinetic theory of heat. Ever since Clausius [4] postulated that "the energy of the universe is constant and the entropy of the universe strives to attain a maximum value" while passing only through thermodynamic equilibrium states, Maxwell [5] asserted that: "One of the best established facts in thermodynamics is that it is impossible in a system enclosed in an envelop which permits neither change of volume nor passage of heat, and in which both the temperature and the pressure are everywhere the same, to produce any inequality of temperature or of pressure without the expenditure of work. This is the second law of thermodynamics, and it is undoubtedly true as long as we can deal with bodies only in mass, and have no power of perceiving or handling the separate molecules of which they are made up …". Then he conceives his omniscient and omnipotent brain child who can contradict a circularly postulated second law of thermodynamics, and concludes … "In dealing with masses of matter, while we do not perceive the individual molecules, we are compelled to adopt what I have described as the statistical method of calculation, and to abandon the strict dynamical method, in which we follow every molecule by the calculus".

Maxwell's sharp-witted being was subsequently nicknamed "Maxwell's intelligent demon" by Thomson [6], and created what Thomson called the reversibility paradox [7], that is, raised the question: "How can irreversibility result from molecular motions and collisions which are themselves (according to Newton's laws of motion) reversible in time?"

Next, Boltzmann tried to explain the second law of thermodynamics, misconstrued as $dS > 0$, by using classical mechanical principles, but concluded that such an explanation could not be completed without the statistical approach introduced by Maxwell [8]. In 1876, Loschmidt brought the reversibility paradox to the attention of Boltzmann and he quickly converted the apparent difficulty into a new conceptual advance. He asserted that "systems tend to pass from ordered to disordered states, rather than the reverse, because the number of disordered states is so much greater than the number of ordered states". Moreover, this explanation suggested to Boltzmann that "entropy – previously a rather mysterious quantity – should be interpreted as a measure of disorder, and he specified the very well known expression etched on his tomb stone $S = k\log\Omega$". The idea of disorder has been adopted by many preeminent scientists, including Feynman [9], Penrose [10], and Denbigh [11].

Statistical theories of thermodynamics yield many correct and practical numerical results only about thermodynamic equilibrium states [12, 13]. Over the past almost two centuries however, despite these successes, thousands of scientists and engineers [14] have expressed a dissatisfaction with the almost universal efforts to compel thermodynamics to conform to statistical explanations in the light of both many accurate, reproducible nonstatistical experiences and many theoretical inconsistencies that have been identified, and a desire for a better theory, as proposed for the first time by Carnot [15]. The adherence to statistical explanations is contrary to the response of the scientific community to the evidence about the heliocentricity of our solar system, and the revolutionary modifications of classical mechanics introduced by the theories of relativity, and quantum mechanics.

Over the past three decades, intrigued and challenged by the prevalent misunderstandings and misconceptions about thermodynamics, a small group at MIT has proposed two intimately interrelated resolutions of the dilemmas and paradoxes created by the statistical interpretations of thermodynamics, in general, and by the Maxwell explanation, in particular. One of the resolutions is purely thermodynamic without reference to quantum theory, and the other quantum thermodynamic without statistical probabilities. Moreover and perhaps more importantly, the advances just cited provide definitive and fully documented answers to the questions raised by Einstein one hundred years ago about the molecular-kinetic theory of heat, and Brownian movement. For the purposes of this article and better communication, the various aspects of the resolutions proposed by the MIT group are not discussed in the chronological order in which they were developed.

**A novel nonquantal exposition of thermodynamics**

Gyftopoulos and Beretta [16] have composed a novel, nonstatistical exposition in which all concepts of thermodynamics are defined completely and without circular and tautological arguments in terms of only the concepts of space, time, and force or inertial mass, plus three noncircularly, unambiguously, and completely defined postulates. Though the intellectual underpinning of this exposition is the unified quantum theory summarized later, what follows does not require any quantum-theoretic concepts, postulates, and theorems. It is noteworthy, however, that the three postulates of thermodynamics turn out to be theorems of the unified quantum theory of mechanics and thermodynamics.

The order of introduction of concepts, postulates, and theorems is: *system* (types and amounts of constituents that can range from one spin to any number of spins and/or other particles, forces between constituents, and external forces or parameters, such as the dimensions that define a volume, and shape of volume); *properties and their values at an instant in time* and not as averages over an infinite period of time; *state* in terms of the concepts just cited; *first law as an assertion that any two states of a system can be initial and final states of a weight process between the system and its environment* (and not in terms of energy, work, and heat that have not yet been defined); definition of *energy* as a theorem of the first law; *energy balance* in terms of energy change of the system and energy exchanged between the system and its environment; *classification of states in terms of evolutions in time such as unsteady, steady, nonequilibrium, equilibrium, and stable – thermodynamic – equilibrium; second law as an assertion of existence of one and only one stable equilibrium state for each set of values of energy, amounts of constituents, and parameters such as volume V* ( and not in terms of entropy and temperature that have not yet been defined)*; definition of a reservoir; generalized available energy as a property of a system combined with a reservoir; entropy as a property of any state of the system (stable equilibrium or not) in terms of energy and generalized available energy*\* (and not in terms of temperature and heat that have not yet been defined); *entropy balance* in terms of entropy change of the system and entropy exchanged with systems in the environment of the system plus entropy generated by irreversibility (*if any*) inside the system.\*\*

A system having an amount of energy denoted by $E$, r different constituents with amounts denoted by $\bm{n} = \{n_1, n_2, ..., n_r\}$, and s different parameters with values denoted by $\bm{\beta} = \{\beta_1, \beta_2, ..., \beta_s\}$, can be in one of an infinite number of states. However, *the second law asserts that one and only one of these states is a globally stable equilibrium state*. It follows that any property of a system in a stable equilibrium state must be a function only of $E$, $\bm{n}$, and $\bm{\beta}$. In particular, the entropy of a stable equilibrium state must be of the form $S(E, \bm{n}, \bm{\beta})$ and this form is called the *fundamental relation*. Among many practical applications, the fundamental relation is used for the definitions of *temperature*, *pressure*, and *total potentials*, properties that are valid only for stable equilibrium states. In particular, for a system that has volume as the only parameter, the definitions of the properties just cited are as follows:

Temperature: $T = (\partial S/\partial E)_{\bm{n}, V}$ (1)

Total potentials for i = 1, 2, ..., r:
$\mu_i = (\partial E/\partial n_i)_{S, \bm{n}} = -T(\partial S/\partial n_i)_{E, \bm{n}, V}$ (2)

Pressure: $p = T(\partial S/\partial V)_{E, \bm{n}}$ (3)

In connection with temperature, the *third law* of thermodynamics asserts that: *for each set of values of the amounts of constituents and the parameters of a system without upper limit on energy, there exists one stable equilibrium state with zero temperature or infinite inverse temperature $1/T = \infty$, and for a system with an upper limit on energy, such as a spin system, there exist two stable equilibrium states with zero temperature, one with $1/T = +\infty$, and the other with $1/T = -\infty$.*

It is noteworthy that $T$ and $\mu_i$'s are interpreted as measures of escaping tendencies in the following sense. Given two systems A and B in stable equilibrium states, energy and entropy flow from A to B if and only if $1/T_A < 1/T_B$ for both positive and negative $T_A$, and constituent i flows and interacts from A to B for $T_A = T_B$ if and only if $(\mu_A)_i > (\mu_B)_i$. Pressure is interpreted as measure of a capturing tendency for volume.

Next, we define the terms work and heat. *Work is an interaction between a system and its environment that involves only the exchange of energy. Heat is an interaction between a system*

---

\*It is noteworthy that entropy is shown to be a property only of the system despite the fact that generalized available energy is a property of both the system and the reservoir.
\*\*It is noteworthy that the laws of thermodynamics do not require that entropy must always increase. Two very important theorems are: (i) if a weight process is reversible, the entropy remains invariant; and (ii) if a weight process is irreversible, the entropy increases. The key word in both theorems is "if".

*and a special system in its environment. The special system is a reservoir that can exchange any amount of energy at the constant temperature of the reservoir and a concomitant amount of entropy equal to the ratio of the energy divided by the fixed temperature of the reservoir.* Other heat interactions are discussed in Ref. [16].

Neither work nor heat is contained in a system. Energy and entropy are properties of any state of a system, whereas work and heat are modes of interactions; said differently, no method exists to identify that in any state, thermodynamic equilibrium or not thermodynamic equilibrium, a fraction x of the energy is work, and the remainder $1 - x$ is heat for any value $0 \leq x \leq 1$. So, whatever the cause of Brownian movement, it is not the molecular-kinetic conception of heat.

Many other important theorems of the novel exposition of thermodynamics are discussed in Ref. [16].

**Neutral stable equilibrium states**

The fundamental relation depends on the volume of a system but not on the shape of the volume. For example, two systems containing identical types and amounts of constituents, one being a cube of volume $V_c$ and the other an orthogonal prism of volume $V_p = V_c$, have identical fundamental relations. It follows that

$$(dS)_{E, n, V} = S_c - S_p = 0 \qquad (4)$$

To be sure the preceding equality is valid for any pair of equal volumes regardless of the shape of each volume.

Systems that satisfy Eq. 4 for any pair of equal volumes have been discussed by Hatsopoulos and Keenan [17], and are said to be in *neutral stable equilibrium*, the neutrality referring to the constancies of the energy, amounts and types of constituents, and the size of the volume regardless of its geometric shape. But we must recognize that changes of shape play a decisive role in determining how the parameters affect the behavior of the system, and as we will see in the theoretical understanding of Brownian movement.

For example, for a cubical shape of side equal to a, the volume $V_c = a^3$ and only the parameter "a" determines the details that enter in the evaluation of the value of the fundamental relation. On the other hand, if the shape is an orthogonal prism with sides b, c, and d, and volume $V_p = bcd = a^3 = V_c$ (Figure 1), then there are three parameters that enter in the evaluation of the fundamental relation. Even though this change of parameters does not affect the value of the fundamental relation, it affects the configuration of the constituents which in the cube is determined only by "a", whereas in the prism it is determined by "b", "c", and "d". We will see later that the change of configuration is continuous in time, and is evidenced as Brownian movement. Another example of changes of shapes is represented by the white background and the black caricatures in Figure 2.

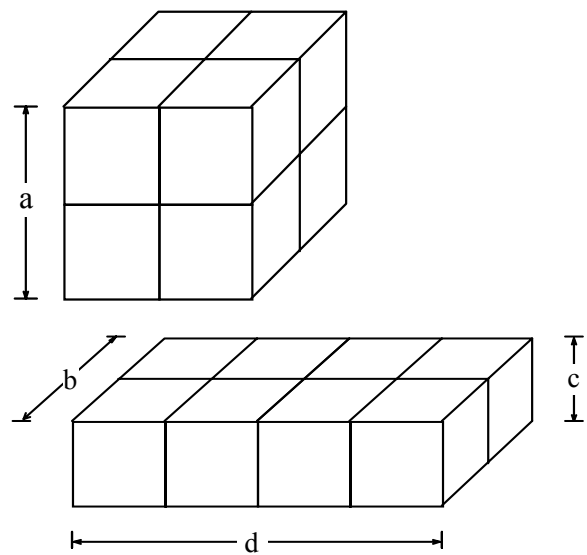

Figure 1: A simple example of shape change without volume change.

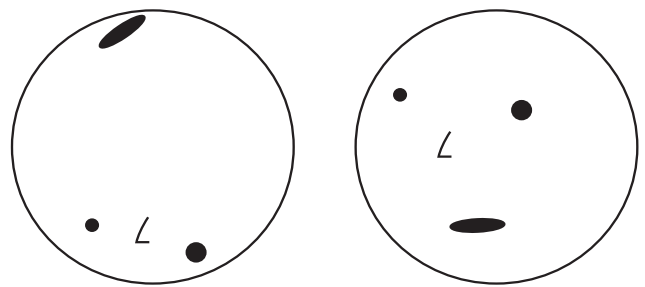

Figure 2: A realistic example of shape changes of a solvent (white) and a colloid (black spots) without changes of the volumes of either the solvent or the colloid.

**Thermodynamic analysis of Brownian movement**

Systems in which Brownian movement is observed are in a neutral thermodynamic equilibrium state and consist of two phases: (i) a *liquid solvent* capable of dissolving and/or dispersing one or more other phases; and (ii) a colloid composed of particles much larger than atoms or ordinary molecules but much too small to be visible to an unaided eye, dispersed but not dissolvable in (i).

For our purposes, we consider an isolated composite system consisting of a liquid solvent A, and a dispersed but not dissolvable colloid B in a neutral stable equilibrium state that has energy $E = E_A + E_B$, amounts of constituents $\mathbf{n} = \mathbf{n}_A + \mathbf{n}_B$, volume $V = V_A + V_B$, temperature T, pressure p, and $E_i$, $\mathbf{n}_i$ and $V_i$ have fixed values for i either A or B. Persistent experimental observations indicate that in such a system the constituents of the colloid and, consequently, also of the liquid solvent appear to aimlessly, and for all practical purposes endlessly be moving around over observed time periods of many years. This is what Einstein and many other scientists call Brownian movement, and there is no doubt whatsoever about the existence of the phenomenon.

In the novel nonquantal exposition of thermodynamics, we have established that for a system with fixed values *E, n, and V* there exists one and only one stable equilibrium state. In a definitive exorcism of Maxwell's demon [18], we prove that in a system in a stable equilibrium state no particle of any constituent is moving – each particle has zero velocity. So, Brownian movement appears to contradict these results. However, the contradiction is illusory for the following reason.

In the exposition of thermodynamics, we find that each total potential is interpreted as an escaping tendency [16]. Moreover, we prove that if a constituent is not present in a system then its total potential is equal to minus infinity [19]. So for the solvent and the colloid the following relations apply for each constituent i of the solvent, and j of the colloid:

$$\left(\mu_{\text{solvent}}\right)_i > \left(\mu_{\text{colloid}}\right)_i = -\infty \tag{5}$$

$$\left(\mu_{\text{colloid}}\right)_j > \left(\mu_{\text{solvent}}\right)_j = -\infty \tag{6}$$

So, systems in which Brownian movement is observed are in partial mutual stable equilibrium, that is, they satisfy the conditions of temperature and pressure equalities but not the conditions of total potential equalities. As a result, both the constituents of the solvent and the colloid exert infinitely large "driving forces" (total potential differences) on the interface between the two phases, and try to interpenetrate each other as they would have done if the colloid were soluble by the solvent. However, such interpenetration is impossible, and the only effect is a continuous in time modification of the pliable shape of the interface between the two phases, the battle goes on forever and appears to any observer as Brownian movement. Said differently, it is not phase motions that cause the observed movements but infinitely large differences in total potentials that change the shape of the interface and appear as motions.

In the quantum thermodynamic discussion of the problem, we provide explicit results of the effects of continuous interfacial shape changes.

**A unified quantum theory of mechanics and thermodynamics**

Hatsopoulos and Gyftopoulos [20] have conceived a nonstatistical resolution of the dilemmas and paradoxes that have preoccupied generations of physicists over more than a century in their attempts to rationalize the relation between mechanics and thermodynamics. The resolution is based on a unified quantum theory of mechanics and thermodynamics which without modification encompasses all systems (both macroscopic and microscopic, including systems of only one particle or one spin), and all states (both stable – thermodynamic – equilibrium, and not stable equilibrium). The key and distinguishing features between statistical expositions of thermodynamics and the unified quantum theory of mechanics and thermodynamics are as follows:

(i) The recognition that the quantum-mechanical density operators $\rho \geq \rho^2$ that are subject to the laws of physics (quantum-theoretic and thermodynamic) are those that can be represented by a *homogeneous ensemble*. In such an ensemble, every member is assigned the same $\rho$ as any other member, and experimentally (in contrast to algebraically), ρ cannot be decomposed – is unambiguous or irreducible – into a statistical mixture of either projectors or density operators different from ρ. Graphical illustrations of homogeneous ensembles, and statistical mixtures or *heterogeneous ensembles* are shown in Figures 3 and 4. The relevance and reality of unambiguous density operators has also been identified by Jauch [21], and had been observed by Schrödinger [22] who, however, did not pursue the consequences of his observation.

(ii) The recognition that the Schrödinger equation of motion is correct but incomplete. It is incomplete because it describes only zero entropy evolutions in time that are unitary and therefore reversible. The same remarks apply to the von Neumann equation of motion for statistical density operators which correspond to nonzero entropy. But not all reversible evolutions in time are unitary and not all evolutions are reversible. In response to this recognition, Beretta et al [23, 24] conceived a nonlinear equation of motion for ρ that has as a limiting case the Schrödinger equation (zero entropy physics) and regularizes evolutions in time that are reversible and either unitary or nonunitary, and evolutions that are irreversible. Definitions, postulates, and major theorems of the unified quantum theory are summarized also in the Appendix of Ref. [25].

(iii) The determination of the analytical expression for entropy [26] which differs from each and every of the dozens of expressions that have been proposed in the literature but is the only one that satisfies nine criteria that have been established in the quantal and nonquantal expositions of thermodynamics. The criteria are as follows. The expression must: (1) be invariant for all unitary evolutions in time; (2) be well defined for every system, both large and small, and for every state, stable equilibrium or not stable equilibrium; (3) be invariant in all reversible adiabatic processes; and increasing in all irreversible processes; (4) be additive for all systems and all states; (5) be nonnegative and vanish for all states encountered in conventional quantum mechanics, namely states for which the probabilities are described by a projector $\rho = \rho^2$; (6) have a unique value for given values of energy, amounts of constituents, and parameters if the system is in a stable equilibrium state; (7) result in a concave and smooth graph of entropy versus energy for stable equilibrium states that correspond to given values of amounts of constituents and parameters; (8) yield the same values of temperature, total potentials, and pressure for a composite system C consisting of two systems A and B in mutual stable equilibrium; and (9) reduce to relations that have been established experimentally and that express the entropy in terms of values of energy, amounts

## HOMOGENEOUS ENSEMBLE

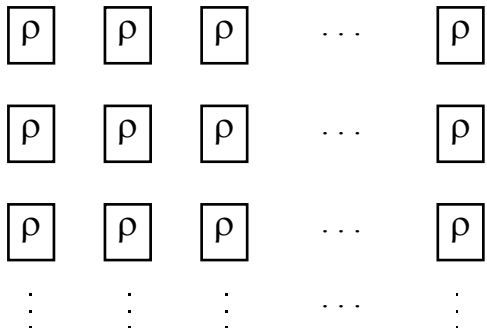

OVERALL DENSITY OPERATOR = ρ

Figure 3: Pictorial representation of a homogeneous ensemble. Each of the members of the ensemble is characterized by the same density operator $\rho \geq \rho^2$. It is clear that any conceivable subensemble of a homogeneous ensemble is characterized by the same ρ as the ensemble.

## HETEROGENEOUS ENSEMBLE

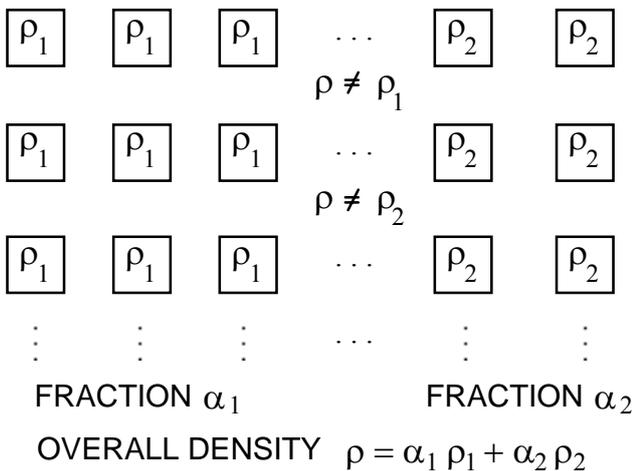

FRACTION $\alpha_1$       FRACTION $\alpha_2$

OVERALL DENSITY $\rho = \alpha_1 \rho_1 + \alpha_2 \rho_2$

Figure 4: Pictorial representation of a heterogeneous ensemble. Each of the subensembles for $\rho_1$ and $\rho_2$ represents either a projector $\left(\rho_i = \rho_i^2\right)$ or a density operator $\left(\rho_i > \rho_i^2\right)$ for i = 1, 2, and $\alpha_1 + \alpha_2 = 1$.

of constituents, and parameters, such as the relations for ideal gases.

Out of the dozens of expressions that have been presented in the archival literature, one and only one satisfies all the criteria just cited and is the relation

$$S = -k\text{Tr}[\rho\ln\rho] \qquad (7)$$

In appearance, this relation is identical with the von Neumann entropy of statistical quantum mechanics in which $\rho$ is a statistical average of projectors. Conceptually, however, the relation differs radically from the von Neumann entropy because here $\rho$ is not a statistical average of projectors, but a quantum mechanical operator represented by a homogeneous ensemble (Figure 3).

(iv) The von Neumann entropy is by definition restricted to apply only to thermodynamic equilibrium states, whereas the entropy of quantum thermodynamics does not have that restriction.

(v) Whereas in deriving an equation of motion for $\rho$ of statistical quantum mechanics the fraction of each projector in the heterogeneous ensemble is simultaneously treated as both time independent and time dependent [27], the density operator of quantum thermodynamics does not have this discrepancy.

(vi) The entropy of quantum thermodynamics is a measure of the spatial shape of the constituents of the system in any state, stable equilibrium or not stable equilibrium [28, 29]. None of the statistical entropies have such a characteristic.

(vii) Whereas the entropies of statistical mechanics are thought to represent ultimate disorder if the system is in a stable equilibrium state, the entropy of the unified theory represents perfect order for such a state [30].

## Quantum thermodynamic analysis of Brownian movement

Consider the same system defined for the thermodynamic analysis of Brownian movement. The constituents of both the liquid solvent and the colloid worm aimlessly and endlessly their ways within each other because of the infinitely large differences between total potentials (Eqs. 5, 6), while each phase is passing through a neutral stable equilibrium state, that is, the energy, amounts of constituents, and volume of each phase is fixed but the parameters are changing in a very complex and sinuous manner. As a result, the energy eigenprojectors and eigenvalues of both the liquid solvent and the colloid and the corresponding stable equilibrium state density operators change continuously in time so as to accommodate the continuously changing shapes - parameters- of the volumes of the two systems of the composite of the liquid solvent and the colloid. These continuous changes in time are impossible to evaluate because both of the lack of knowledge of the precise change of the shapes of the volumes, and the difficulty inherent in calculating eigenprojectors and eigenvalues in cases of complicated shapes of even very simple systems such as one particle in an odd looking, one dimensional potential well. The laws of physics, however, have no difficulty in continuously in time responding to the changing shapes of the liquid solvent and the colloid and determining the nonstatistical density operators for each pair of shapes at each instant in time.

An illustration of the quantum mechanical effects of the changes of the shapes of the volumes of a composite system consisting of a solvent and a colloid in neutral stable equilibrium states has been made by Çubukçu [31]. He considers the Hamiltonian operators $H_g$ for $g = s \text{ or } c$, where s is the solvent, and c the colloid, and proceeds with the following calculations:

Hamiltonian operator of the two systems

$$H = H_s \otimes I_c + I_s \otimes H_c \qquad (8)$$

where I is the identity operator

Energy eigenfunctions and eigenvalues

$$H_s\psi_i = e_i\psi_i \qquad H_c\varphi_j = \varepsilon_j\varphi_j \qquad (9)$$

$$H(\psi_i \otimes \varphi_j) = (e_i + \varepsilon_j)(\psi_i \otimes \varphi_j) \qquad (10)$$

Density operator of both phases

$$\rho(t) = \rho_s(t) \otimes \rho_c(t) \qquad (11)$$

Eigenvalues of density operator

$$\rho(t)(\psi_i \otimes \varphi_j) = p_{ij}(\psi_i \otimes \varphi_j) \qquad (12)$$

Relation between density operator eigenvalues and energy eigenvalues

$$\ln(p_{ij}/p_{kl}) = \left[(e_k - e_i) + (\varepsilon_l - \varepsilon_j)\right]/kT \qquad (13)$$

for all pairs of indices $\{i, j\}$ and $\{k, l\}$, where $T$ is the constant temperature of the solvent and the colloid.

From Eq. 13 we see that as the eigenvalues and eigenfunctions of the solvent and the colloid change, the constituents of each of the two phases are continuously reallocated to the evolving energy eigenstates, and the reallocation appears as Brownian movement. Said differently, the reallocations are the cause of the motions, and not the motions the cause of the reallocations.

## Brownian motion as understood by *e. coli*

At a special meeting of the New England Sections of APS and AAPT at MIT, April 1-2, 2005, celebrating the 100[th] anniversary of the miraculous year of Einstein's contributions to science, one of the keynote presentations was on "Brownian Motion, as understood by *E. coli*" and was delivered by Professor Berg. In his book "*E. coli* in Motion" [32], Berg describes the physics of *E. coli* as follows: "The physics that looms large in the life of *E. coli* is not the physics that we encounter, because we are massive and live on land, while *E. coli* is microscopic and lives in water. To *E. coli*, water appears as a fine-grained substance of inexhaustible extent, whose component particles are in continuous riotous motion. When a cell swims, it drags some of these molecules along with it, causing the surrounding fluid to shear. Momentum transfer between adjacent layers of fluid is very efficient, and to a small organism with very little mass, the viscous drag that results is overwhelming. As a result, *E. coli* is utterly unable to coast: it knows nothing about inertia. When you put in the numbers you find that if a cell swimming 30 diameters per second were to put in the clutch, it would coast less than a tenth of the diameter of a hydrogen atom! And a tethered cell spinning 10Hz would continue to rotate for less than a millionth of a revolution. But cells do not actually stop, because of thermal agitation. Collisions with surrounding water molecules drive the cell body this way and that, powering brownian motion. For a swimming cell, the cumulative effect of this motion over a period of 1 second is displacement in a randomly chosen direction by about 1 $\mu$m and rotation about a randomly chosen axis by about 30 degrees. As a consequence, *E. coli* cannot swim in straight line. After about 10 seconds, it drifts off course by more than 90 degrees, and thus forgets where it is going. This sets an upper limit on the time available for a cell to decide whether life is getting better or worse. If it cannot decide within about 10 seconds, it is too late. A lower limit is set by the time required for the cell to count enough molecules of attractant or repellent to determine their concentrations with adequate precision. The number of receptors require for this task proves surprisingly small, because the random motion of molecules to be sensed enables them to sample different points on the cell surface with great efficiency."

Though the description is accurate, our view is that it is not related in any way whatsoever to Brownian motion for the following reasons.

In contrast to systems in which Brownian movement is observed, and which consist of a liquid solvent and a colloid, each maintaining its identity for all practical purposes for ever, *E. coli* have a totally different biography. Relevant statements that one finds in textbooks on molecular biology of the gene [33] and biochemistry [34] are as follows: "… the bacteria *E. coli* will grow in an aqueous solution containing just glucose and several inorganic ions" … "There is a lower limit, however, to the time necessary for a cell generation; no matter how favourable the growth conditions, an *E. coli* is unable to divide more than once every 20 minutes. … The average *E. coli* cell is rod shaped. … It grows by increasing in length followed by a fission process that generates two cells of equal length." In addition, *E. coli* is self propelled not as a result of infinite differences between total potentials such as exist between a liquid solvent and a colloid but because of flagella.

In view of all these facts, one must conclude that the time dependent changes of *E. coli* studied by molecular biologists, and biochemists do not represent Brownian motions. They are the result of chemical reactions between *E. coli* and the nutrients supplied by the solvent.

## Concluding remarks

Though there is no way to prove it or disprove it, I have a strong feeling that the man of the 20$^{th}$ century would have been happy with the thermodynamic discussion of Brownian movement because he was a great admirer of even the traditional version of the subject. He expressed his admiration by the following statement: "A theory is the more impressive the greater the simplicity of its premises, the more different kinds of things it relates, and the more extended its area of applicability [35]." Of course, I must add that he might be very unhappy with the quantum thermodynamic part of this essay because of the serious doubts he expressed about quantum theory in his Einstein, Podolsky, Rosen paper [36] and his exchange of views with Schrödinger [37-39]. Finally, inherent in my exposition of thermodynamics is the idea that entropy has nothing to do with the so-called arrow of time (see footnote about irreversibility), and Einstein would fully agree with this observation because in his last letter to his very close friend Besso he wrote: "For us true physicists, the distinction between past, present, and future times is an illusion, tenacious as it may be [40]."